\documentclass[12pt]{iopart}

\usepackage{graphicx}
\usepackage{epstopdf}
\usepackage{iopams}  
\usepackage[latin1]{inputenc}
\usepackage[T1]{fontenc}
\usepackage[english]{babel} 
\usepackage{array}
\usepackage{lmodern}
\usepackage{siunitx}
\usepackage[parfill]{parskip}
\usepackage{multirow}

\begin{document}

\title{Towards all-optical atom chips based on optical waveguides}

\author{Yuri B. Ovchinnikov, Folly Eli Ayi-Yovo}

\address{National Physical Laboratory, Hampton Road, Teddington TW11 0LW, UK}
\ead{yuri.ovchinnikov@npl.co.uk}
\vspace{10pt}
\begin{indented}
\item[]4 December 2019
\end{indented}

\begin{abstract}
Coherent guiding of atoms in two-colour evanescent light fields of two main single modes of suspended optical rib waveguides is investigated theoretically. Special attention is paid to waveguides of widths larger than the wavelength of light, which provide better lateral stability of the surface traps and waveguides, and can be used in coherent Bragg beam splitters for matter waves, based on optical gratings formed by interference of evanescent light waves of two crossed optical waveguides. A single-mode regime for evanescent-wave waveguides for atoms is investigated. The general structure and key elements of all-optical atom chips are discussed.

\end{abstract}

%
%
%
%
%

\section{Introduction}

All-optical atom chips for coherent control of ultra-cold atoms, based on conservative optical dipole forces, potentially have several advantages over existing magnetic atom chips, which are based on integrated current-carrying wires or permanent magnets \cite{1}. The sizes of individual all-optical elements of such chips, like traps, waveguides and beam splitters, can be comparable to the wavelength of light, providing a clear path to miniaturisation of the corresponding atom chips. The optical dipole traps can confine atoms in magnetically insensitive internal states, which makes these chips attractive to applications in which the influence of environmental magnetic fields is unwanted. Finally, such chips ideally should not dissipate too much heat, and therefore should not need additional cooling.

In this paper we will consider only the conservative trapping and guiding of atoms in non-resonant two evanescent light fields of different colour and penetration depth \cite{2}.

There were already several proposals made to create surface waveguides for atoms \cite{3,4,5,6,7} based on two-colour evanescent light waves generated above dielectric ridge optical waveguides of subwavelength transverse sizes. All previous attempts to realize guiding or trapping of atoms near planar waveguides were not successful so far, which we attribute to bad optical quality of the optical waveguides or difficulties of loading of atoms into the surface traps \cite{7}. On the other hand, a stable trapping of Cs atoms at the distance of about 200 nm from the surface of a dielectric optical nanofibre has been successfully demonstrated \cite{8,9}.

In the last several years there has been rapid development of suspended optical waveguides based on silicon \cite{10} or silica \cite{11} thin membranes. One of the main reasons for the high interest in suspended waveguides is an enlarged penetration depth of optical evanescent waves into surrounding medium of the waveguide, which is beneficial for using in different types of sensors based on evanescent light waves. Due to equally high contrast of the refractive indexes at both interfaces between the core of suspended planar waveguide and surrounding medium, which is vacuum in our case, it can support modes with large propagation angles, which correspond to the largest penetration depths of the surrounding evanescent waves. This is not the case for planar waveguides on top of a dielectric substrates, where the maximum propagation angle of the guided mode is limited by the contrast of refraction indexes at the interface between core of the waveguide and substrate.

In this paper we consider the use of all-optical waveguides for trapping and guiding of ultra-cold atoms in a field of two-colour evanescent light waves generated by two single modes of suspended SiO$_2$ rib waveguides supported by a silicon substrate. 

Special attention is paid to waveguides which are several times wider than the wavelength of guiding light. There are several advantages of such wide planar optical waveguides for guiding and trapping of atoms. First, the weak lateral optical confinement of such wide rib waveguides provides very small optical losses of the propagating modes at the mutual intersection of the waveguides, which makes it possible to build new integrated atom optics elements, like traps,
two-dimensional arrays of traps, continuous beam splitters for matter waves and
complicated guiding circuits for atoms. Second, the intersection of the rib
waveguides with each other can generate an optical evanescent lattice in the
cross region, which can be used as a coherent waveguide Bragg beam splitter for
matter waves \cite{12,13}. Third, the cross-section of such wide evanescent wave
surface waveguides for matter waves is essentially larger than the cross-section
of sub-wavelength atomic waveguides based on ridge optical waveguides \cite{3,4,5,6,7},
which leads to a proportional increase in the maximal atomic flux through these
waveguides. The high atom flux is very important to increase the
signal-to-noise ratio of the corresponding integrated waveguide atom
interferometers. Fourth, as it is shown in this paper, such waveguides make it possible to provide stable guiding of atoms in a single-mode regime, which is very important for realisation of waveguide atom interferometers.

The paper is organised as follows. In the second section the stability conditions for atomic traps based on two-colour evanescent light fields are discussed. The third section presents numerical modelling of suspended rib optical waveguides, which support only fundamental modes of the laser light of
two different colours, and gives the main parameters of the corresponding
evanescent light fields above the rib, which are used for trapping of rubidium
atoms. In the fourth section the results of calculations of the corresponding optical dipole potentials and their basic parameters are presented.  In the fifth section a structure and some key elements 
of all-optical atom chips are discussed. Finally, in the conclusion we summarise
the results of our investigations.

\section{Stability conditions for atomic traps and waveguides on evanescent waves}

\subsection{Potential of the trap}

The two-dimensional atomic trap (waveguide) consists of two evanescent waves (EW), formed in vacuum due to total internal reflection of light from a planar dielectric surface (see Fig.1). We assume that the two waves have different frequencies, such as one of them is red-detuned and the other is blue-detuned from the main electron transition of a certain atom. The red-detuned evanescent wave forms negative (attractive) dipole potential for the atom, while the blue-detuned evanescent wave forms a positive (repulsive) dipole potential. The corresponding penetration depths of the evanescent light waves in vacuum are $d_r$ and $d_b$. Note, that in optical waveguides the penetration depth of evanescent fields into a clad (vacuum) is proportional to the wavelength of light and $d_r>d_b$. The lateral spatial distributions of the two evanescent light waves are described by functions $A_{r}(y)$ and $A_{b}(y)$. We are considering here only one lateral dimension of the trap along y-axis, which can be easily generalised to any 3D evanescent wave trap.  

Initially we will neglect the presence of additional Van der Waals interaction between the atoms and the dielectric substrate. In that case the corresponding total dipole potential of the trap is

\begin{equation} \label{EQ1} 
U\left(x,y\right)=U_re^{\left(-2{x}/{d_r}\right)}A_r{\left({y}\right)}+U_be^{\left(-2{x}/{d_b}\right)}A_b{\left(y\right)},
\end{equation} 

where $U_r<0$ and $U_b>0$.

\subsection{Potential along surface normal}

First we consider the transverse trapping at the centre of the trap ($y = 0$), where the potential along $x$-axis is simplified to

\begin{equation} \label{EQ2} 
U\left(x,0\right)={U}_r\left(x\right)+U_b\left(x\right)=U_re^{\left(-2{x}/{d_r}\right)}+U_be^{\left(-2{x}/{d_b}\right)} 
\end{equation} 

The minimum of the potential, which is determined by the equation

\begin{equation} \label{EQ3} 
{dU\left(x,0\right)}/{dx}=0,      
\end{equation} 

is reached at the distance

\begin{equation} \label{EQ4} 
x_0=\frac{1}{2}\frac{d_r d_b}{d_r-d_b}{\mathrm{log} \left(-\frac{d_r U_b}{d_b U_r}\right)\ }.      
\end{equation} 

From the equation (\ref{EQ4}) it follows that the trapping outside the border of the dielectric sudstrate is possible only if $d_r > d_b$. 

It is easy to derive from (\ref{EQ4}) that the ratio between the amplitudes of red-detuned and blue-detuned potentials at the equilibrium point is

\begin{equation} \label{EQ5} 
\frac{\left|U_r\left(x_0\right)\right|}{\left|U_b\left(x_0\right)\right|}=\frac{d_r}{d_b}. 
\end{equation} 

The above condition can be achieved only if the potentials at the border of the dielectric substrate satisfy to the inequality

\begin{equation} \label{EQ6} 
\frac{\left|U_r\right|}{\left|U_b\right|}<\frac{d_r}{d_b}. 
\end{equation}
  
Otherwise, the dipole potential has no minimum outside the substrate. 

\subsection{Lateral confinement}

The lateral distribution of the total dipole potential of the two-colour EW trap in vicinity of the equilibrium point ($x=x_0$, $y=0$) can be approximated as

\begin{equation} \label{EQ7} 
U\left(x_0,y\to{0}\right)=U_r\left(x_0\right)\left(1-y^2/q_r^2\right)+U_b\left(x_0\right)\left(1-y^2/q_b^2\right), 
\end{equation}
for values $q_r$, $q_b$ describing behaviour of $A_r(y)$, $A_b(y)$.

 In that case, the potential (\ref{EQ7}) has a minimum near $y=0$ point only when the amplitude of attractive force of red-detuned field grows faster than the amplitude of repulsive force of blue-detuned field near the equilibrium point, where both of these radial forces are equal to zero. Mathematically it corresponds to the next ratio

\begin{equation} \label{EQ8} 
\left|\frac{d^2\left(U_r\left(x_0\right))\left(1-y^2/q_r^2\right)\right)}{dy^2}\right| >\left|\frac{d^2\left(U_b\left(x_0\right)\left(1-y^2/q_b^2\right)\right)}{dy^2}\right|.    
\end{equation} 

For small values of $y$ the inequality (\ref{EQ8}) is reduced to 

\begin{equation} \label{EQ9} 
\left|\frac{2U_r\left(x_0\right)}{q^2_r}\right|>\left|\frac{2U_b\left(x_0\right)}{q^2_b}\right|. 
\end{equation} 

Taking into account the ratio (\ref{EQ5}) we can get from (\ref{EQ8}) the next condition for the lateral confinement of the EW trap 

\begin{equation} \label{EQ10} 
\frac{d_r}{d_b}>\frac{q^2_r}{q^2_b} .    
\end{equation} 

If the condition (\ref{EQ10}) is not fulfilled, the atoms are not trapped in the radial direction, but expelled from the trap. Based on that inequality it is convenient to introduce a stability factor

\begin{equation} \label{EQ11} 
\tau =\frac{d_r q^2_b}{d_b q^2_r} .    
\end{equation} 

The EW trap is stable in the lateral direction when $\tau >1$ and unstable otherwise.

\subsection{EW waveguide for atoms above an optical planar waveguide}

In a case of an optical strip-like (strip, ridge, rib) planar waveguide, which is directed along the $z$-axis, the problem is reduced to 2D-trapping along the $x$ and $y$ axes. The transverse lateral intensity distribution of light in the core region of such a waveguide can be rather well approximated by 

\begin{equation} \label{EQ12} 
A_{r,b}(y)={{\mathrm{cos}^2 \left(y/q_{r,b}\right)\ }}. 
\end{equation} 

Close to the equilibrium point, $A_{r,b}(y)$ follows the same harmonic approximation as in (7), so the condition for lateral stability of the trap (11) in the absence of the Van der Waals potential can be applied.
According to the physics of propagation of the guided modes of the waveguide, the evanescent field of the red-detuned mode, which has larger wavelength, has a larger penetration depth in surrounding vacuum, compared to the blue-detuned mode. This leads (Fig. 1) to larger lateral size of the red-detuned EW compared to the blue-detuned EW. As a result of that, the lateral stability of the trap is not always fulfilled and depends on exact parameters of the trap, including the Van der Waals interaction of atoms with the dielectric substrate, which is considered in the next section. 

The attractive interaction between atoms and dielectric surface of the substrate can be approximated with the next potential

\begin{equation} \label{EQ13} 
U_{VdW}(x)=-\frac{C_3 \lambda_{eff}/(2\pi)}{x^3(x+\lambda_{eff}/(2\pi))}\ , 
\end{equation} 

where $C_3$ is the Van der Waals coefficient and $\lambda_{eff}$ is the reduced atomic transition wavelength, which is providing transition from the Van der Waals interaction at $x<<\lambda_{eff}/(2\pi)$ to the Casimir-Polder interaction at $x>>\lambda_{eff}/(2\pi)$. It is valid to use such a formula as far as a typical distance of the EW trap from the substrate is comparable to  $\lambda_{eff}$, where the Casimir-Polder potential become dominant.  

\begin{figure}
	\centering
	\includegraphics[scale=0.5]{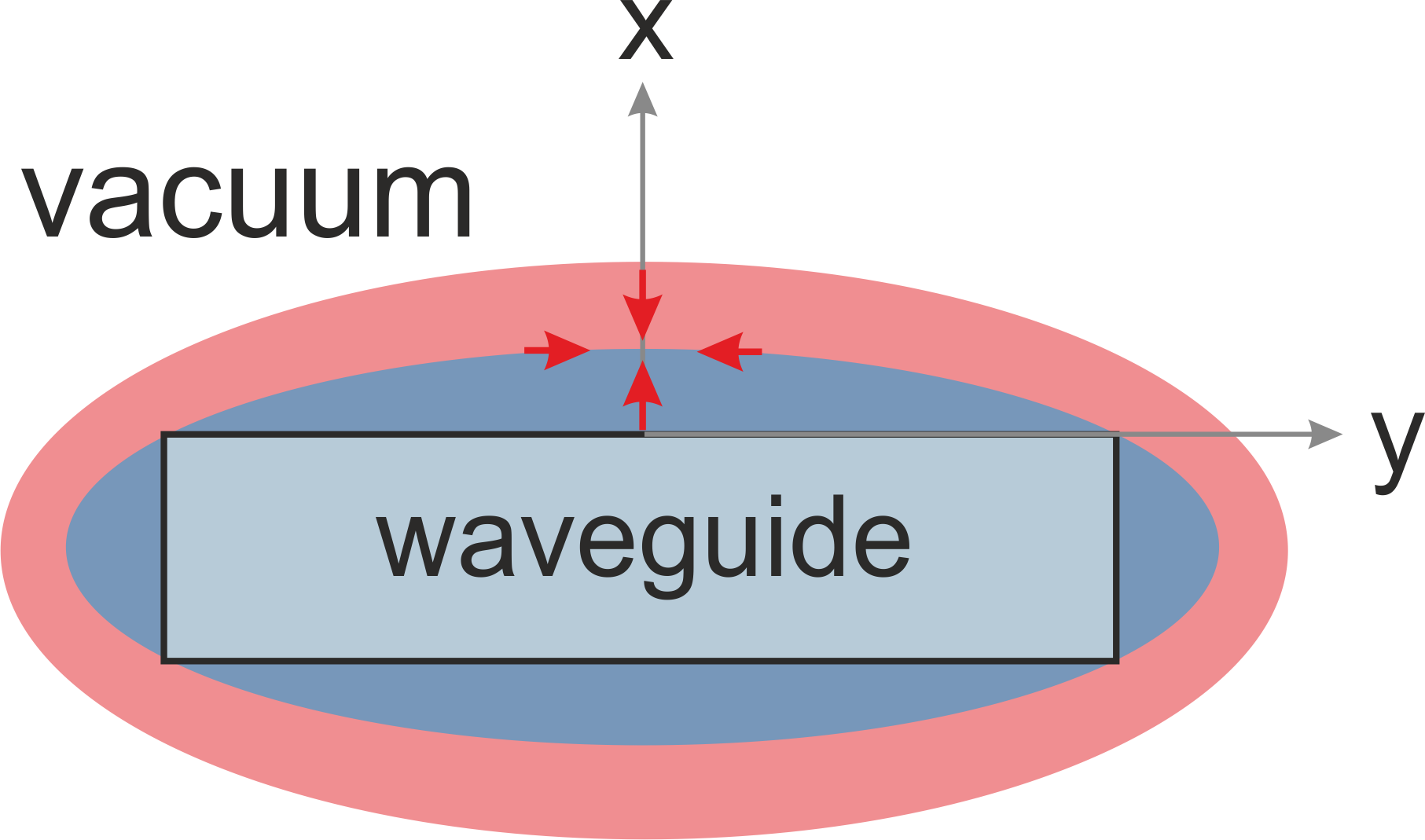}
	\label{}
	\caption{Evanescent light fields near rectangular dielectric waveguides, which supports propagation of two main optical modes of different colour.}
\end{figure}

 In the presence of the attractive Van der Waals force, the position of the minimum of the dipole trap shifts along the x axis from the equilibrium point of the dipole trap $x_0$ to a new position $x_0 - \Delta x$. To find that displacement $\Delta x$ we first approximate the dipole potential near the initial equilibrium point $x_0$ as a harmonic one. The corresponding dipole force near that point is $F_{dip}=-kx$. It is easy to show that the spring constant in that force is

\begin{equation} \label{EQ14} 
k=4U_b\left(x_0\right)\frac{d_r-d_b}{d_rd^2_b}\ . 
\end{equation} 

In the case of small displacement $\mathrm{\Delta }x\ll d_r,d_b$, the equilibrium displacement is equal to

\begin{equation} \label{EQ15} 
\mathrm{\Delta }x=\frac{F_{VdW}(x_0)}{k}\ , 
\end{equation} 

where $F_{VdW}(x_0)=\partial U_{VdW}(x_0)/ \partial x$ is the Van der Waals force near the equilibrium point. The value of the displacement (\ref{EQ15}) is derived under the assumption that the Van der Waals force is constant in the vicinity of the minimum of the trap, which is a valid approximation for $x_0 \sim \lambda_{eff}$.

This displacement of the trap minimum also changes the lateral stability condition for the EW waveguide (or trap). It is possible to show that the corresponding stability factor in presence of the Van der Waals interaction becomes

\begin{equation} \label{EQ16} 
{\tau }^*=\tau \frac{1+{\mathrm{2}\mathrm{\Delta }x}/{d_r}}{1+2{\mathrm{\Delta }x}/{d_b}} .    
\end{equation} 

Due to the fact that in a stable EW trap $d_r>d_b$, the stability factor ${\tau }^*$ is always less than $\tau $.

\section{Planar optical rib waveguides for two-colour fields}


\subsection{Rib optical waveguide approach}

\begin{figure}[h]
		\centering
		\includegraphics[scale=0.5]{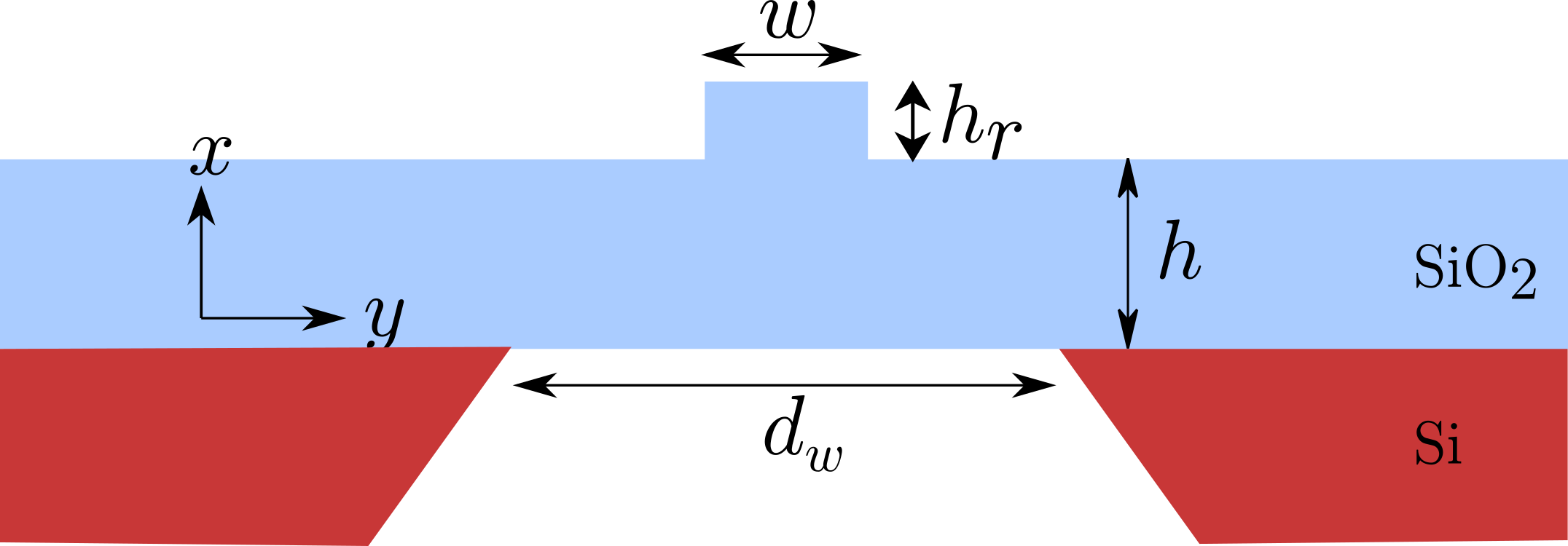}
		\caption{Cross section of a rib optical waveguide with simulation parameters}
	    \label{fig_rib_section}
\end{figure}

In order to realise the wide integrated waveguide for matter waves based on two-colour evanescent light waves, suspended rib optical waveguides \cite{10,11} can be used. The waveguide geometry is shown in Figure \ref{fig_rib_section}. The rib width is $w$ = 2 $\mu$m. The top and bottom cladding consist of air. The core material is silicon dioxide (SiO$_2$). For the wavelengths, we have chosen 720 nm and 850 nm. $h$ and $h_r$ are thicknesses of the slab and the rib respectively. A typical transverse size of the membrane is $d_w\geq$ 40 \si{\micro\meter}. The decay lengths and the lateral mode sizes of the waveguide fundamental modes will be used to compute the trap. Therefore, optical simulations were carried out to design single mode suspended rib waveguide for SiO$_2$ platforms. 

\subsection{Rib optical waveguide design}

We use numerical simulations to identify geometrical parameters for single mode operation for both wavelengths and for TE polarisation. The refractive index of SiO$_2$ is 1.45. The single mode limits for $h_r$ were computed as a function of $h$.

The properties of rib waveguides have been studied extensively. Soref et al \cite{14} first proposed an analytical formula to determine the single mode condition (SMC).

\begin{equation} \label{EQ17} 
\frac{h}{H}\leq 0.5;
\end{equation} 

\begin{equation} \label{EQ18} 
\frac{w}{H} = \alpha+ \frac{\frac{h}{H}}{\sqrt{1-(\frac{h}{H})^2}}   
\end{equation} 

where  $\alpha$=0.3 and $H=h+ h_r$. Other approaches were also proposed in order to address this issue. 
Pogossian et al. \cite{15} modified this formula and suggested a more stringent condition ($\alpha$=0). Xu et al. \cite{16} proposed a solution based on the effective index for silicon-on-insulator (SOI) waveguides. The rib waveguide mode is guided if and only if the effective index of the rib waveguide mode is larger than the effective index of the 1D fundamental TE mode of the outer slab. This approach was also used by Du et al. \cite{17} but on a silicon carbide platform. Dai and Sheng \cite{18} proposed that the mode is guided if the leakage losses are less than 200dB/cm and more than 5\% of the power of the mode is in the rib region. 

Recently, Hui et al. \cite{19} carried out a new study on single mode rib waveguide in silicon photonics. They used three different rules to determine the SMC. The first one is based on the effective index approach. A mode is guided if and only if its effective index is larger than the fundamental mode of the outer slab region of the same polarisation. The outer slab 1D fundamental mode effective index was computed for both TM and TE polarisations. Every TE (respectively TM) mode of the rib region is guided when its effective index is larger than the outer slab 1D effective index of the fundamental TE mode (respectively TM). For the second approach they used the filling factor. If more than 5\% of the power of the mode is confined in the rib region, the mode is guided.  The third approach was based on the propagation losses of the mode. If the mode propagation loss is less than 36 dB/cm the mode is guided. Similar results were obtained using the three approaches. 
  
Rib waveguide with large cross section can be defined as single mode as shown by \cite{10,11}. In theory, large Rib waveguides support higher order modes and they are multimode by definition. However, the losses of the higher-order modes are expected to be high, such that after one or two millimeters of propagation only the fundamental mode remains. Lousteau et al. \cite{20} and recently Leblanc-Hotte et al. \cite{21} show that equations (1) and (2) are not sufficient to ensure the SMC behaviour. The higher order modes losses depend on the geometry and they suggested a case by case study for the design of rib waveguide. 

\begin{figure}[h]
		\centering
		\includegraphics[scale=0.6]{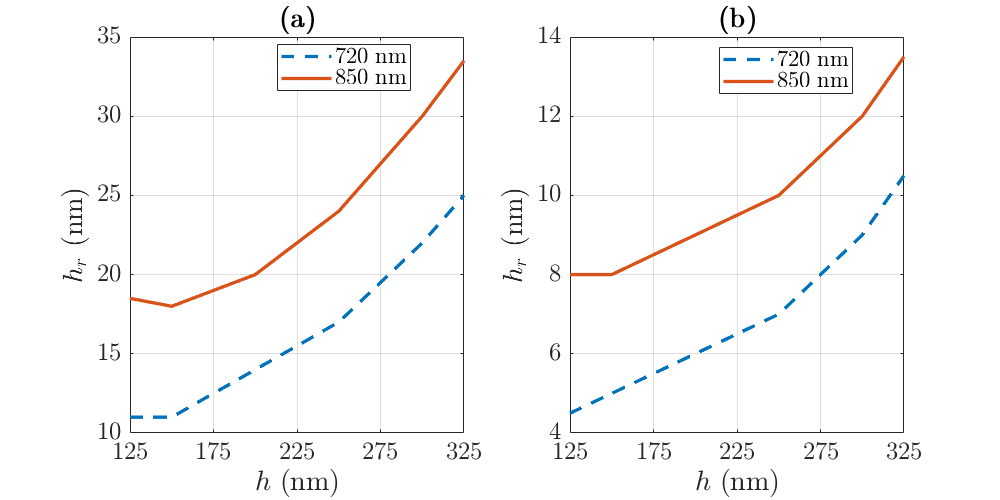}
		\label{fig_rib_SMC}
         \caption{Single mode operation limits as a function of $h_r$ for TE  polarisations and two wavelengths of 720 nm and 850 nm. a) $w$=2 $\mu$m; b) $w$=3 $\mu$m.}
\end{figure}

During this process, as predicted by \cite{20,21}, we noticed that higher order modes can propagate while the SMC based on the effective index method and the one suggested in \cite{19} are fulfilled. We also noticed that the guiding condition of higher order modes depends on the refractive indices and $H$. Indeed, for each value of the refractive index, there is a maximum value of $H_{max}$ for which higher order modes do not exist. For the membrane $H_{max}$=325 nm at 720 nm. Figure \ref{fig_rib_SMC} show the results for TE polarisation for 720 nm and 850 nm for $w$=2 $\mu$m and $w$=3 $\mu$m. The waveguide is single mode below the curve and multimode above. The single mode limit increases with $h$ and the wavelength whereas it decreases with $w$.

The numerical simulations gave us the SMC. Then for each configuration the decay length and the lateral width of the evanescent field is extracted and used to compute the trap potential. 

\subsection{Decay lengths and lateral widths of the evanescent light waves}

The extraction of the decay length and the lateral mode size is described in this section. This will be performed for $w$=2 $\mu$m $h$ = 300 nm and $h_r$ = 15 nm. This choice takes into account the manufacturing constraints and the stability of the trap. On one side, a smaller slab thickness will lead to a smaller rib thickness which will be difficult to fabricate. On the other side, a larger slab thickness will lead to higher order modes which would affect the stability of the trap. This configuration is a compromise between both constraints and also provides good penetration depth which decreases the power in the waveguide. 

\begin{figure}[h]
		\centering
		\includegraphics[scale=0.6]{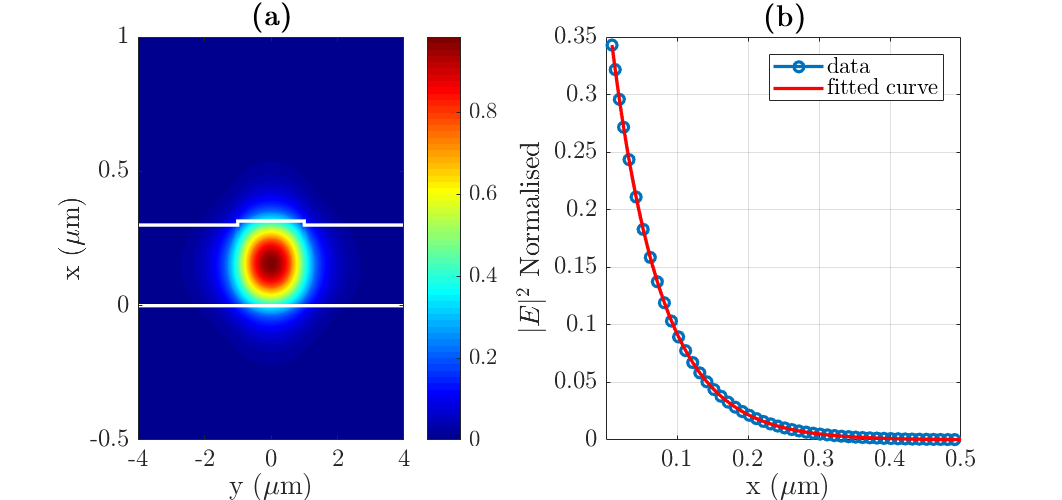}
		\caption{Profile of a fundamental mode of a rib waveguide (a) and intensity on top of the waveguide at $y$=0 (b)}
         \label{fig_decay_length}
\end{figure}

The electromagnetic (EM) field intensity $I$ ($I\propto \ {\left|E\right|}^2)$ distribution of the fundamental TE mode was computed. Then, the modal field intensity was extracted along $\left(Ox\right)$ axis for $y$= 0. Finally, the decay length was computed by interpolating the evanescent part of the EM field on the top of the waveguide. Figure \ref{fig_decay_length}(a) shows the EM field intensity distribution over the cross section of the rib waveguide.

For the calculation of the decay length, the evanescent part of the EM field intensity (on the top of the rib waveguide) was interpolated into the form of $I(x)=\ I_0 e^{-2x/d}$ where $d$ is the decay length and $I_0$ the maximum intensity on the top of the rib waveguide. Figure \ref{fig_decay_length}(b) shows the decaying part of the intensity $I$ on the top of the rib waveguide (the positions on $\left(Ox\right)$ axis are counted from the top of the rib waveguide) for simulations and after interpolation.

The lateral intensity distribution of the fundamental TE mode above the rib region is:

\begin{equation} \label{EQ19} 
I\left(y\right)\sim {{\mathrm{cos}^2 \left(y/q\right)\ }} 
\end{equation} 

Near the central region of the rib on the top of the waveguide, the intensity is described by the following relation:

\begin{equation} \label{EQ20} 
I\left(y\right)=I_0\left(1-\frac{y^2}{q^2}\right) 
\end{equation} 

where $q$ is the lateral mode size of the waveguide and $I_0$ the maximum intensity along $\left(0y\right)$ on the top of the rib waveguide.

\begin{figure}[h]
		\centering
		\includegraphics[scale=0.6]{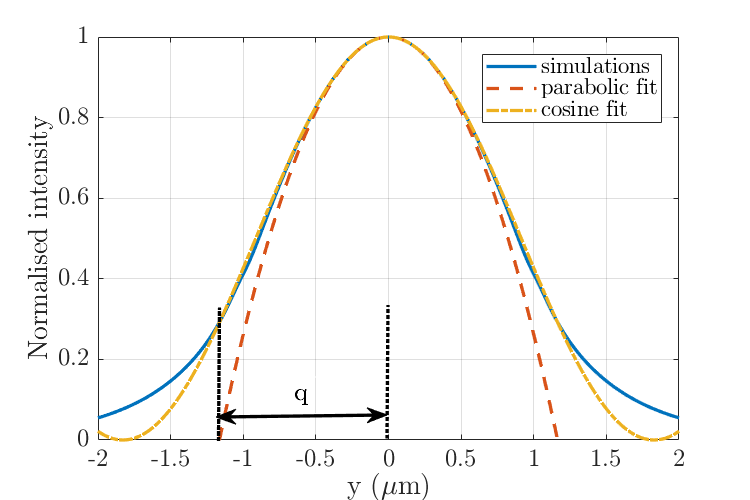}
		\caption{Mode intensity on top of the rib waveguide from simulations and interpolations.}
         \label{fig_lateral_size}
\end{figure}

Figure \ref{fig_lateral_size} show the fundamental TE mode intensity for 720 nm on the top of the waveguide, the different interpolations (cosine and parabolic) using equations \ref{EQ17} and \ref{EQ18}. 

Table \ref{table_trap_params} shows the decays lengths and the lateral mode sizes at 720 nm and 850 nm for the fundamental TE mode for some configurations.

\begin{table}
\begin{center}
\caption{ Decay lengths and lateral mode sizes for the fundamental TE mode for 720 nm and 850 nm}
\begin{tabular}{|p{1in}|p{1in}|p{1in}|p{1in}|}
	\hline
	\multicolumn{4}{|c|}{$w$=2 \si{\micro\meter},  $h$ =300 \si{\nano\meter}, $h_r$= 15 \si{\nano\meter} } \\
	\hline
	\multicolumn{2}{|c|}{$\lambda$ = 720 nm} & 	\multicolumn{2}{c|}{$\lambda$ = 850 \si{\nano\meter}} \\
	\hline
	$d_b$ (\si{\nano\meter}) & 139.5 & $d_r$ (\si{\nano\meter}) & 175.0\\
	\hline
	$q_b$ (\si{\micro\meter}) & 1.164 & $q_r$ (\si{\micro\meter}) & 1.435 \\
	\hline
\end{tabular}
\end{center}
\label{table_trap_params}
\end{table}

\section{Transverse trapping potential of EW atomic waveguide}

\subsection{Trapping potential along surface normal}

We will consider first the trapping potential along $x$ axis at the centre of waveguide ($y$=0), which includes interaction of atoms with a dielectric substrate. It is described by the formula

\begin{equation} \label{EQ21} 
U\left(x,y\right)=U_{r\ }{{\mathrm{exp} \left(-{2x}/{d_r}\right)}}+U_{b\ }{{\mathrm{exp} \left(-{2x}/{d_b}\right)-\frac{C_3{{\lambda }_{eff}}/{2\pi }}{x^3\left(x+{{\lambda }_{eff}}/{2\pi }\right)}\ }\ },  
\end{equation} 

where the first two terms correspond to the optical dipole potentials of the two evanescent wave fields and the third term represents the potential of interaction between rubidium atoms and dielectric surface of SiO$_2$ rib waveguide, $C_3$ = 5.699 $\times 10^{-49}$ Jm$^3$ is the van der Waals coefficient, $\lambda_{eff}$ = 710 nm \cite{21}. The corresponding total potential for Rb atoms in optical evanescent fields of two main modes with wavelengths $\lambda_r$=850 nm and $\lambda_b$=720 nm, penetration depths $d_r$=175.0 nm and $d_b$=139.5 nm and light intensities at the dielectric surface $I_r$=6.5$\times$10$^{9}$ W/m$^2$ and $I_b$=1.2$\times$10$^{10}$ W/m$^2$, which correspond to the rib optical waveguide of width $w$=2 $\mu$m, $h=300$ nm and $h_r$=15 nm, is presented in Fig.6a. The potential is characterised by the potential well depth $U_{min}$ and the maximum value of the potential barrier $U_{max}$, which separates the well from dielectric surface. The value of the maximum of potential barrier can be positive or negative, depending on parameters of the two evanescent waves. On Fig.6a and Fig.6b it is shown that an increase of the intensity of red-frequency-detuned EW at fixed intensity of the blue-frequency-detuned EW leads to shift of trap minimum towards dielectric surface and a corresponding decrease of the height of potential barrier due to the dominant influence of the attractive Van der Waals potential at distances $x \lesssim$ 100 nm. 

Fig.7a shows the dependence of values $U_{min}$ (lower curve) and $U_{max}$ (upper curve) on the intensity of the red-detuned EW at the surface of dielectric substrate at fixed intensity of the blue-detuned EW $I_b$=1.2$\times$10$^{10}$ W/m$^2$. The decay lengths of the evanescent waves in these calculations were equal to $d_r$=175 nm and $d_b$=139.5 nm. Increase of intensity of the red-detuned field first leads to an increase of the trap depth, which is equal to $U_{min}$ at $U_{max} \geqslant 0$. It leads also to a shift of the trap minimum towards the surface of the substrate ($x$=0), as is shown in Fig. 7b. For higher intensities of the red-detuned field, when the trap minimum shifts below $\simeq$200 nm, where influence of the Van der Waals attraction becomes dominant, the value of $U_{max}$ becomes negative and the total depth of the trap becomes equal to  $U_{min}-U_{max}$, until it turns to zero at $U_{min}=U_{max}$. Certainly the best conditions for stable trapping of atoms are achieved at  $U_{max} > 0$, when the trap has full depth of $U_{min}$ and is separated from the wall by a high enough potential barrier, which prevents quantum tunnelling of atoms to the surface of dielectric substrate. 
 
\begin{figure}
	\centering
	\includegraphics[scale=0.4]{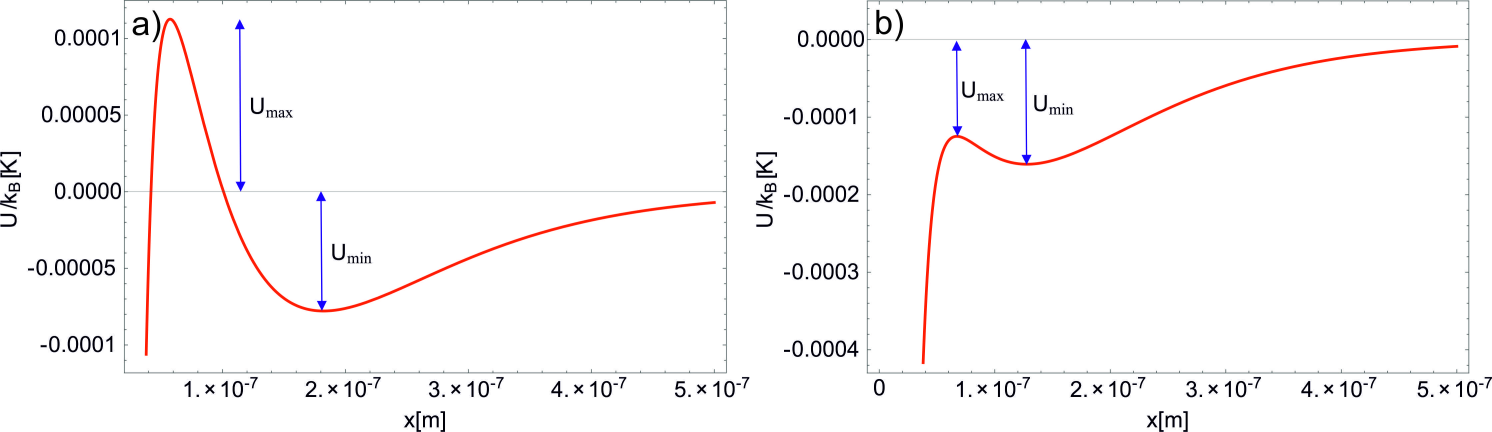}
	\label{}
	\caption{a) Spatial distribution of trapping potential along $x$-axis for Rb atoms at light intensities $I_r$=6.5$\times$10$^{9}$ W/m$^2$ and $I_b$=1.2$\times$10$^{10}$ W/m$^2$; b) The same distribution at $I_r$=7.5$\times$10$^{9}$ W/m$^2$ and $I_b$=1.2$\times$10$^{10}$ W/m$^2$.}
\end{figure}

\begin{figure}
	\centering
	\includegraphics[scale=0.4]{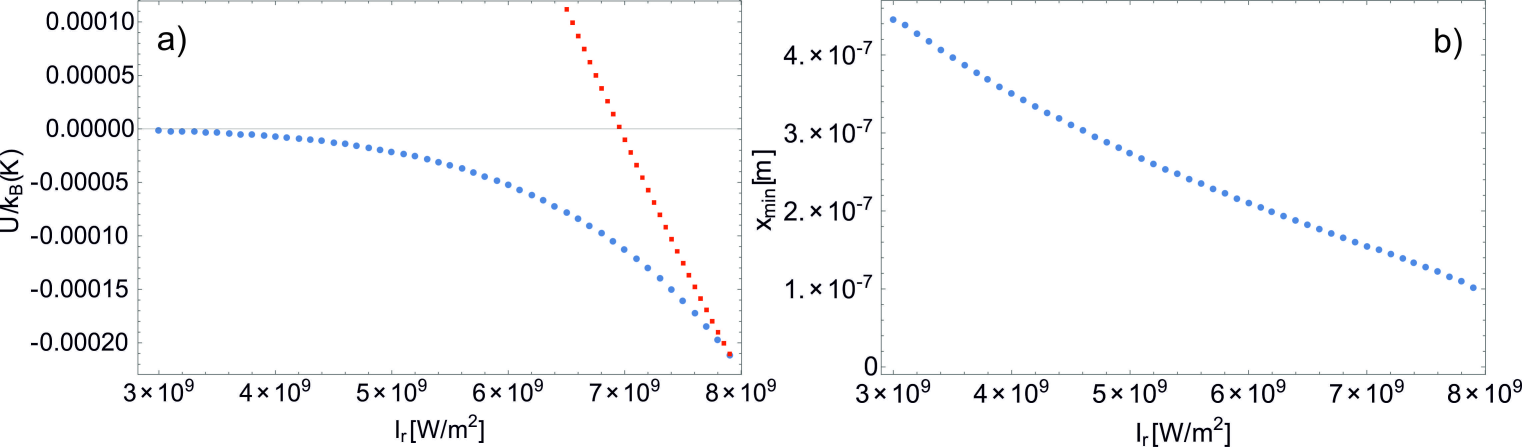}
	\label{}
	\caption{a) Dependence of $U_{min}$ (lower curve) and $U_{max}$ (upper curve) on the intensity $I_r$ of the red-detuned at fixed intensity of the blue-detuned field $I_b$=1.2$\times$10$^{10}$ W/m$^2$; b) Dependence of the position of the trap minimum on $I_r$ for the same fixed $I_b$.}
\end{figure}

Similar dependence of $U_{min}$ (lower curve) and $U_{max}$ (upper curve) on the intensity of blue-detuned field $I_b$ at fixed intensity of red-detuned field, $I_r$=5$\times$10$^{9}$ W/m$^2$, is shown in Fig.8a. In that case, an increase of the blue-detuned field moves the trap minimum away from the surface (Fig.8b), while the trap depth $U_{min}$ is decreasing and the height of the potential barrier $U_{max}$ is getting larger.

\begin{figure}
	\centering
	\includegraphics[scale=0.4]{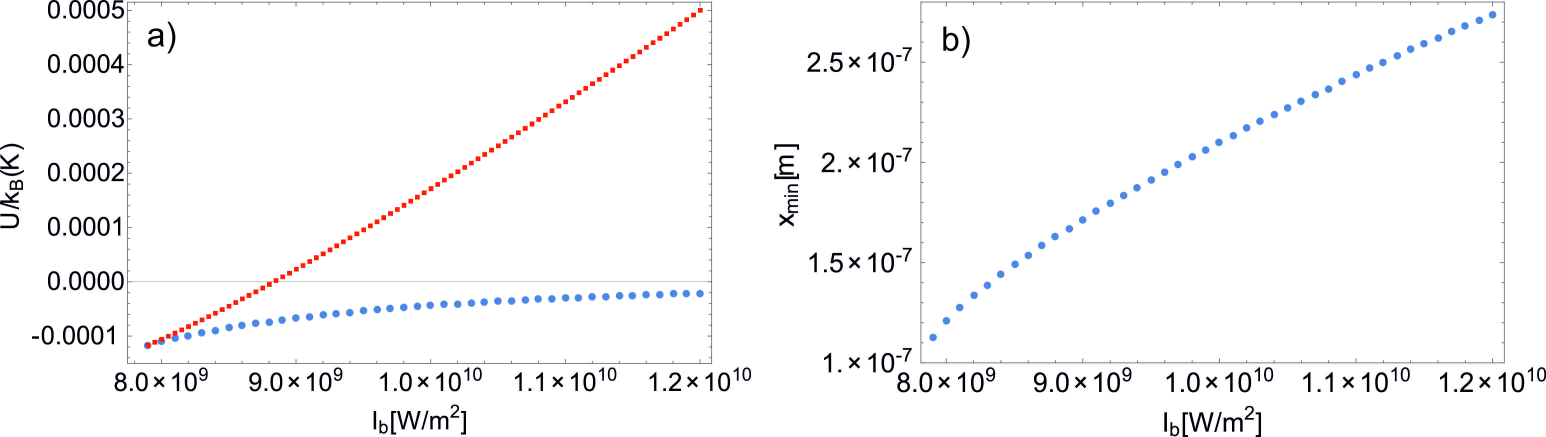}
	\label{}
	\caption{a) Dependence of $U_{min}$ (lower curve) and $U_{max}$ (upper curve) on the intensity $I_b$ of the blue-detuned at fixed intensity of the red-detuned field of $I_r$=5$\times$10$^{9}$ W/m$^2$; b) Dependence of the position of the trap minimum on $I_b$ for the same fixed $I_r$.}
\end{figure}

\begin{figure}
	\centering
	\includegraphics[scale=0.5]{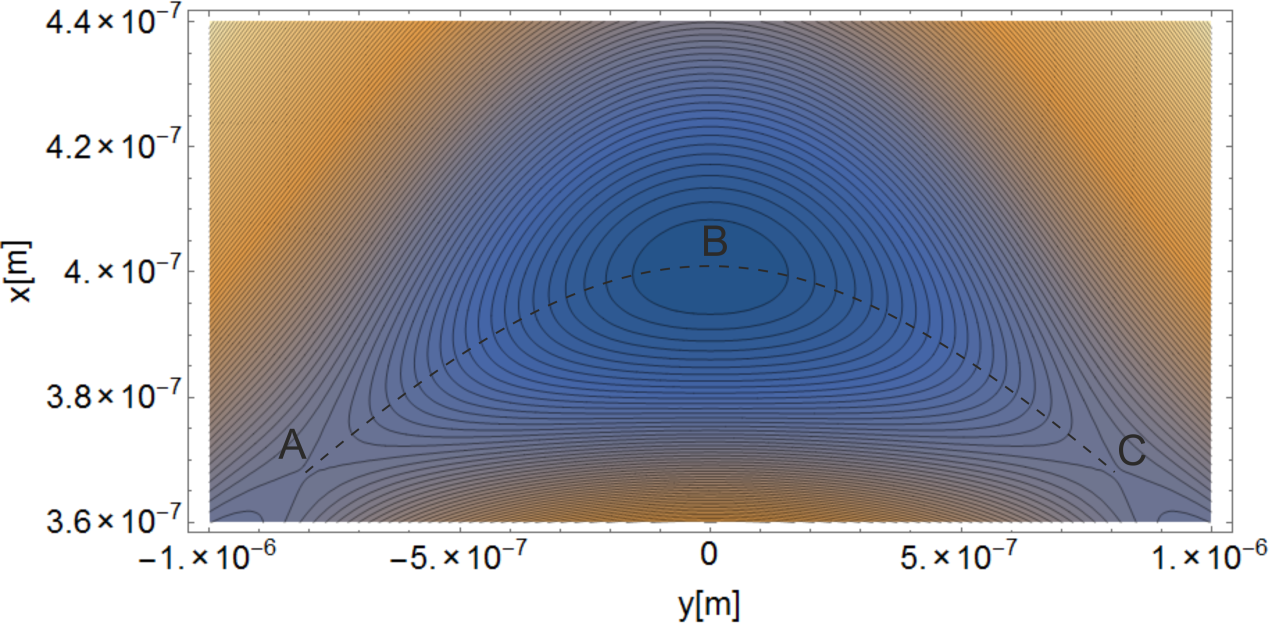}
	\label{}
	\caption{Spatial distribution of the trapping potential in the XY plane for Rb atoms at light intensities $I_r$=1.72$\times$10$^{9}$ W/m$^2$ and $I_b$=6.0$\times$10$^{9}$ W/m$^2$. The dashed line shows the weakest direction of the trap. Change of the background colour from blue to yellow corresponds to increase of the potential.}
\end{figure}

\begin{figure}
	\centering
	\includegraphics[scale=0.45]{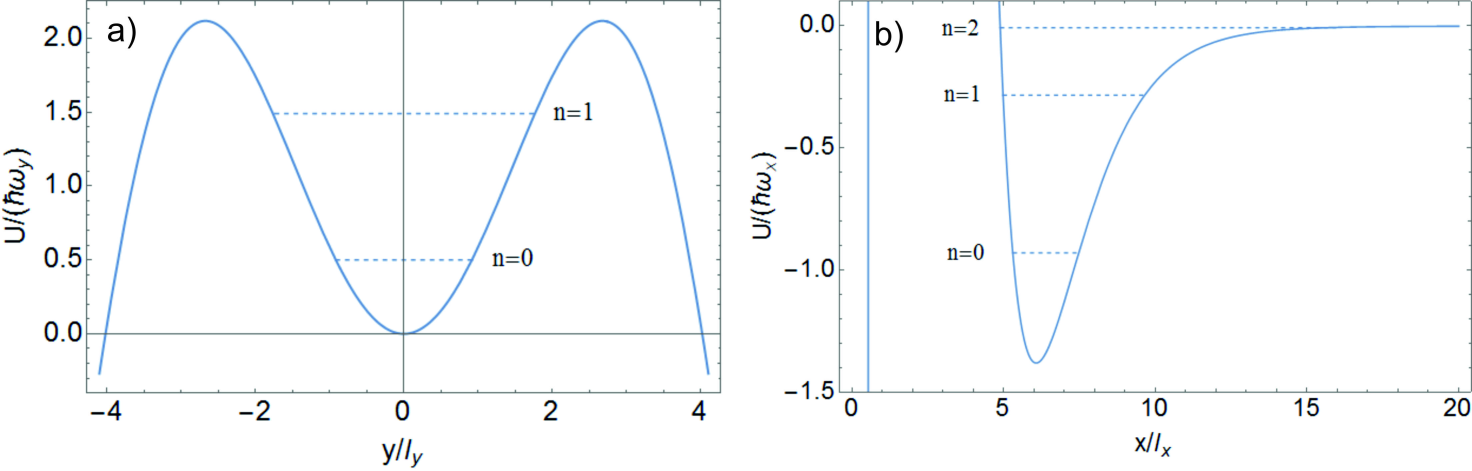}
	\label{}
	\caption{a) Dimensionless spatial distribution of the potential along the dashed line of Fig.9 and its bound states; b) Distribution of the same potential along the $x$ axis at $y=0$ and its three bound states.}
\end{figure}

\subsection{Two-dimensional transverse trapping potential}

To calculate the transverse two-dimensional distribution of the trapping potential in XY-plane we used the formula  

\begin{equation} \label{EQ22} 
U\left(x,y\right)=U_{r}\ {\mathrm{exp} \left(-{2x}/{d_r}\right) A_r\left(y\right)+U_{b\ }{{\mathrm{exp} \left(-{2x}/{d_b}\right) A_b\left(y\right)-\frac{C_3{{\lambda }_{eff}}/{2\pi }}{x^3\left(x+{{\lambda }_{eff}}/{2\pi }\right)}}}}-mgx,
\end{equation} 

where the normalised lateral distributions of the potential $A_r\left(y\right)$ and $A_b\left(y\right)$ were derived from numerical solutions for lateral intensity distributions in evanescent waves of the two fundamental modes of different frequencies inside the corresponding rib waveguide. The last term represents a gravity force directed along the outward normal vector to the substrate. Although, in most cases, the gravity force did not produce much effect on the total trapping potential. Amplitudes of the optical dipole potentials $U_r$ and $U_b$ at $x=0$, $y=0$, which are proportional to the intensity of corresponding light fields, were calculated while taking into account the first eight optical transitions of the principal series of the Rb atom and linear polarisation of light \cite{22}.

First we calculate a shallow transverse trapping potential for a suspended rib waveguide of width $w=2$ $\mu$m, thickness $h=300$ nm and rib height $h_r=15$ nm for the fundamental mode at $\lambda_r=850$ nm and $\lambda_r=720$ nm, with corresponding penetration depths of the evanescent waves of $d_r=175$ nm and $d_b=139.5$ nm. The intensities of light on top of the rib ($x=0, y=0$) are equal to  $I_r$=1.72$\times$10$^{9}$ W/m$^2$ and $I_b$=6.0$\times$10$^{9}$ W/m$^2$, which corresponds to total laser powers in the two modes of $P_r=2.9$ mW and $P_b=9.4$ mW. The exact lateral distributions of intensities in evanescent waves $A_r(y)$ and $A_b(y)$ are taken from numerical simulations of the two modes, while the corresponding parabolic or cosine fits are characterised by lateral lengths $q_r$=1.256 $\mu$m and $q_b$=1.164 $\mu$m. 
The lateral trap stability parameter (\ref{EQ11}) in that case is $\tau=1.077$, which means that the trap is stable in the lateral direction.

The weakest direction of the trap is along the dashed line of Fig.9. The potential depth at point B ($y_0=0, x_0=400.6$~nm) is -1.815 $\mu$K, while the potential depth at the saddle points A and C ($y_0=\pm 0.82 \mu$m, $x_0=367$~nm) is -1.684 $\mu$K. Therefore, the lateral depth of the trap is only 131 nK, while its depth along the $x$-axis is 1.1815 $\mu$K. Such a big difference between lateral and normal depths of the trap is caused by a larger width of the red-detuned evanescent wave compared to the blue-detuned wave, which is an intrinsic property of evanescent light waves around any strip-like optical waveguide (Fig.1).

Corresponding vibrational frequencies of rubidium atom near the bottom of the trap are $\omega_y/2\pi =1.27\ $kHz and $\omega_x/2\pi =26.74\ $kHz. The cross section of the potential along $y$-axis in the centre of trap, along the dashed line, is shown in Fig.10a. The potential is given in units $\hbar \omega_y$, while the coordinate is measured in units $l_y=\sqrt{\hbar/(M \omega_y)}$. The calculated energies of the bound states of the trap are represented by dashed lines. From both of these states an atom can tunnel through the side potential barriers in both directions. The tunnelling rate is calculated to be $\Gamma_{tun}(n=0)$=1.52 s$^{-1}$, which corresponds to a tunnelling probability in 0.1 s of $P_{tun}(t=0.1s)$=0.14. The tunnelling rate from the first excited vibration state is equal to $\Gamma_{tun}(n=1)$=177.2 s$^{-1}$, which makes this state short-lived.
The dimensionless spatial distribution of the potential along the $x$ axis at $y$=0 is shown in Fig.10b. There are three bound states there, the tunnelling rates from which towards the dielectric wall are negligible.
The other factor, which is limiting the coherence time of the trapped or guided atoms in such a potential is a spontaneous scattering rate of photons, which is related to the excitation of higher internal states of the Rb atom. Corresponding scattering rates for the two evanescent fields near the trap minimum ($x_0$=400.6 nm) can be estimated from \cite{22} and are equal to $\Gamma_{scat}$($\lambda$=720 nm)=0.168 s$^{-1}$ and $\Gamma_{scat}$($\lambda$=850 nm)=0.21 s$^{-1}$. Such low scattering rates are achieved due to the small depth of the trap and its large distance from the surface of the substrate. Therefore, at these parameters of the trapping potential it is realistic to achieve coherence times of $t_{coh}\simeq$0.1 s.   

For a deeper trap it is possible to increase the trapping time due to the decrease of lateral tunnelling rates of atoms. For example, we can take the same parameters of the previous trap (Fig.9), except of the intensity of red-detuned evanescent field, which is now increased to $I_r$=1.9$\times$10$^{9}$ W/m$^2$. This increases the total depth of the trap to $U_{min}/k_B=-2.9$ $\mu$K, and the lateral depth to 216 nK, while the trap minimum shifts to $x_0=366.6$ nm. It decreases the lateral tunnelling rate to $\Gamma_{tun}(n=0)$=0.1 s$^{-1}$, while increases the spontaneous scattering rates to $\Gamma_{scat}$($\lambda$=720 nm)=0.27 s$^{-1}$ and $\Gamma_{scat}$($\lambda$=850 nm)=0.34 s$^{-1}$.
Therefore, the total probability of scattering of  $P_{scat}(t=0.15s)$=0.1 can be achieved at $t$=0.15 s.

\subsection{Single mode waveguide for atoms}

A problem of single-mode guiding of atoms is closely related to atom lasers \cite{23} based on Bose-Einstein condensation (BEC) in dilute gases \cite{24} and especially to guided atomic lasers \cite{25,26,27}. All these experiments are using horizontal multimode optical waveguides for atoms based on non-resonant Gaussian laser beams, because in the presence of gravity these beams can't hold atoms in a single-mode regime of the waveguide. Nevertheless, in works \cite{26,27} it was shown that adiabatic loading of the BEC in such a multimode waveguide can preferably load the lowest mode of the waveguide with population of that mode close to 0.85.

Let us derive the conditions at which a waveguide for atoms become single mode waveguide. For simplicity, we consider a harmonic potential $U(x)=\frac{1}{2} M \omega^2 x^2$ of finite depth $U_0$, as it is shown in Fig.11. Such a potential well contains only one bound state if $U_0<\frac{3}{2}\hbar \omega$. This condition for an optical dipole potential can be presented in a form

\begin{equation} \label{EQ23} 
\frac{U_0}{E_R}<\left(\frac{3}{\pi}\right)^2 \frac{\lambda^2}{w_p^2},
\end{equation} 

where $E_r=\hbar^2 k^2/(2m)$ is one photon recoil energy for a light wave with wavevector $k=2\pi/\lambda$, $\lambda$ is the light wavelength and $w_p$ is a total width of the potential at its top. This expression shows that for a large width of the potential $w_p\gg\lambda$ a single mode regime can be achieved only at very small potential depth $U_0\ll E_R$. This makes it clear that only for the integrated optical waveguides for atoms with $w\sim w_p \sim \lambda$ a single mode regime can be reached at $U_0 \sim E_R$. According to equation \ref{EQ23}, it is easier to achieve a single mode regime at smaller width of the waveguide $w$. Nevertheless, our calculations performed for waveguides with width $w=1$ $\mu$m show that the lateral tunnelling rates in such waveguides are so high, that it is impossible to provide stable trapping of atoms in them, unless their potential depth is so high ($\sim$ 45 $\mu$K), that they become not single mode. The depth of the trapping potential for a $w$=2 $\mu$m waveguide considered above (Fig.9 and Fig.10) also can't be further decreased because of the high lateral tunnelling rate of atoms. This tunnelling rate can be decreased only by further increase of the lateral width of the potential and corresponding widths of lateral potential barriers from both sides of the trap.       

\begin{figure}
	\centering
	\includegraphics[scale=0.45]{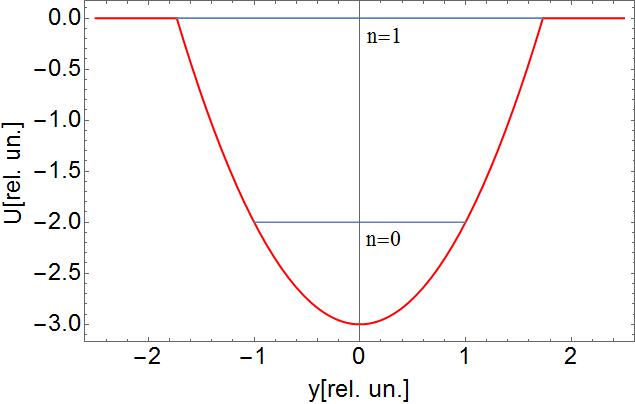}
	\label{}
	\caption{Single mode waveguide for atoms based on harmonic potential of finite depth.}
\end{figure}

The transverse distribution of a very shallow potential for waveguide with $w$=3 $\mu$m, $h=300$ nm and $h_r$=7 nm at light intensities $I_r$=1.3$\times$10$^{9}$ W/m$^2$ and $I_b$=6.0$\times$10$^{9}$ W/m$^2$ and penetration depths of the evanescent waves of $d_r=176.9$ nm and $d_b=140.7$ nm is shown in Fig.12. The total powers in the two modes of the waveguide are $P_r=4.1$ mW and $P_b=13.8$ mW. The lateral trap stability parameter is equal to $\tau$=1.082, which is comparable to the previously considered waveguide with width $w=2$ $\mu$m. The potential minimum in this case is equal to $U_{min}(x=496.9$  nm, $y=0)/k_B=-550.3$ nK, while the lateral depth of the trap is only 44 nK. The dimensionless distributions of the potential along the y- and x-axis with corresponding bound states are shown in Fig.13. Along the $x$ axis only one bound state exists (Fig.13b). The probability of tunnelling from this state towards the dielectric surface is negligibly small. In lateral direction there are two bound states (Fig.13a), the tunnelling rates from which are equal to  $\Gamma_{tun}(n=0)$=0.84 s$^{-1}$ and $\Gamma_{tun}(n=1)$=77 s$^{-1}$ correspondingly. Therefore, the n=1 state can be considered as weakly bound. In that sense this waveguide can be considered as a single-mode waveguide for atoms. In that shallow potential the guided atom experiences maximum confining force along the x-axis is about $10^{24} M g$, while along y-axis the maximum confining force is only $0.5Mg$, where $M$ is the mass of a Rb atom. The spontaneous scattering rates of Rb atoms in this optical potential are estimated to be $\Gamma_{scat}$($\lambda$=720 nm)=0.045 s$^{-1}$ and $\Gamma_{scat}$($\lambda$=850 nm)=0.042 s$^{-1}$.

To increase the coherence time of guided atoms one has to use larger frequency detunings of the guiding light fields from the main optical transition of Rb. For example, we have calculated a waveguide with $\lambda_r=640$ nm and $\lambda_b=930$ nm. Parameters of the rib waveguide are $h=300$ nm $h_r=15$ nm and $w=2$ $\mu$m. Calculated penetration depths of the two evanescent waves in that case are $d_r=199.7$ nm and $d_r=119.9$ nm. The lateral trap stability parameter of $\tau$=1.198 is higher than in previously considered cases, which we attribute to larger frequency difference between the two optical modes. For intensities $I_r$=1.2$\times$10$^{9}$ W/m$^2$ and $I_b$=1.2$\times$10$^{10}$ W/m$^2$ ($P_r=2.1$ mW, $P_b=18.7$ mW) the trap minimum is located at $x_0=328.7$ nm from the surface, where the potential is $-4.48$ $\mu$K, while its lateral depth is $250$ nK. The lateral tunnelling rate from the ground vibrational state is  $\Gamma_{tun}(n=0)$=0.068 s$^{-1}$, while spontaneous scattering rates $\Gamma_{scat}$($\lambda$=640 nm)=0.063 s$^{-1}$ and $\Gamma_{scat}$($\lambda$=930 nm)=0.13 s$^{-1}$. Therefore, in such a waveguide both trapping and coherence times can be about one second.

\begin{figure}
	\centering
	\includegraphics[scale=0.5]{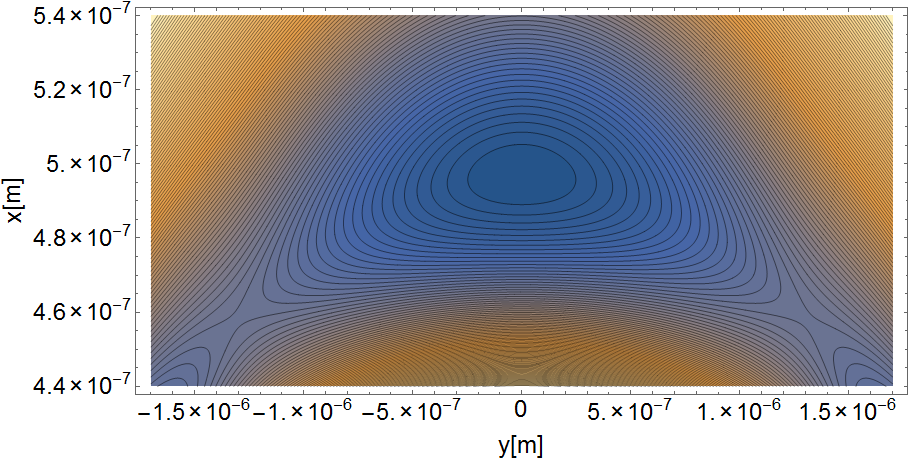}
	\label{}
	\caption{Spatial distribution of trapping potential in XY-plane for rib waveguide of width $w=3\,\mu$m at light intensities $I_r$=1.3$\times$10$^{9}$ W/m$^2$ and  $I_b$=6.0$\times$10$^{9}$ W/m$^2$.}
\end{figure}

\begin{figure}
	\centering
	\includegraphics[scale=0.45]{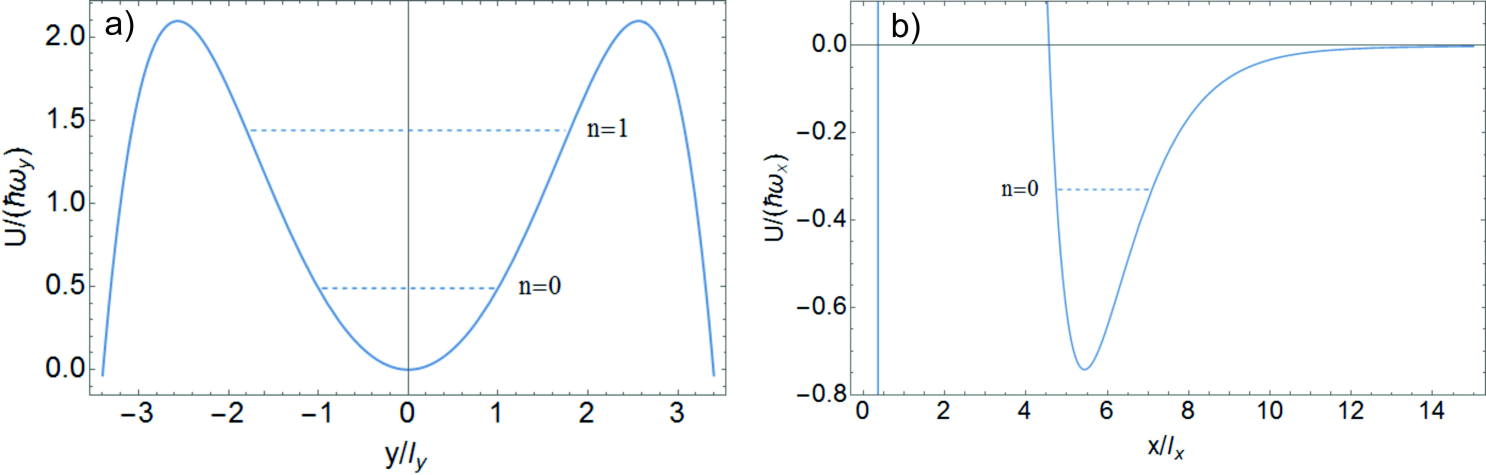}
	\label{}
	\caption{a) Lateral spatial distribution of the potential presented in Fig.12 and its bound states; b) Distribution of the same potential along x-axis at y=0 and its bound state.}
\end{figure}

\section{Atom optics elements for atom chips based on rib waveguides}

\subsection{General structure of all-optical atom chips}

Existing atom chips for ultra-cold atoms are based on magnetic waveguides \cite{1}. All-optical atom chips based on evanescent light waves have a number of advantages over the magnetic chips based on current-carrying wires. First, using optical dipole potentials for trapping and guiding of atoms makes it possible to use atoms in magnetically insensitive internal states, which essentially decrease perturbations of the atomic wave function by fluctuations of the environmental magnetic fields. Second, spatial dimensions of main atom-optics elements of all-optical atom chips can be of the order of several micrometres, which is two orders of magnitude smaller than in a case of magnetic atom chips. The last feature is very important for the miniaturisation of integrated atomic circuits. Third, the light fields can be used for diffraction of atoms, which is essential for realising coherent beam splitters. Finally, the control of light fields is faster and easier than control of magnetic fields.

In general, all-optical atom chips provide a flexible platform for atomic traps, waveguides of different shapes, waveguide resonators for atoms, coherent beam splitters and atom interferometers of different types based on above elements.

\begin{figure}
	\centering
	\includegraphics[scale=0.6]{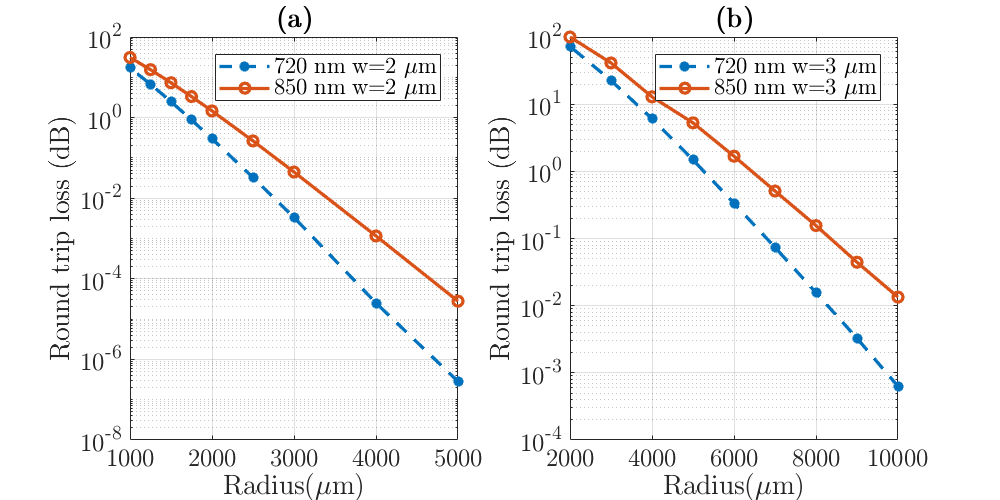}
	\caption{Round trip losses at 720 nm and 850 nm for $w$=2 $\mu$m, $h$=300 nm and $h_r$=15 nm (a) and $w$=3 $\mu$m, $h$=300 nm and $h_r$=7 nm (b) ring waveguides.}
	\label{bend_losses}
\end{figure}

\subsection{All-optical waveguide beam splitters for atoms}

There are two types of waveguide beam splitters, which can be used for coherent splitting of atoms between two integrated waveguides.

First type is a waveguide beam splitter based on optical dual-channel directional couplers \cite{28,29}, which are broadly used in integrated photonic devices. The optical directional coupler consists of two optical waveguides, which are placed very close to each other, such that their propagating modes are partially overlapped with each other. In such a coupler the propagating light can tunnel from one waveguide to the other one. 3-dB splitters based on such directional couplers are used in optical Mach-Zehnder and Sagnac interferometers \cite{30}. Using two-colour light in two coupled rib waveguides can provide a directional coupler for matter waves, which propagate in such all-optical waveguides for atoms. For coherent beam splitter of that type it is preferable to get single mode surface waveguides for matter waves. On the other hand, the main problem in using such directional couplers for atom interferometry is that the corresponding interferometer will be simultaneously an optical and a matter interferometer, and these different types of interferometers can have mutual influence on each other, which might cause additional problems for precise measurements.

\begin{figure}
	\centering
	\includegraphics[scale=0.4]{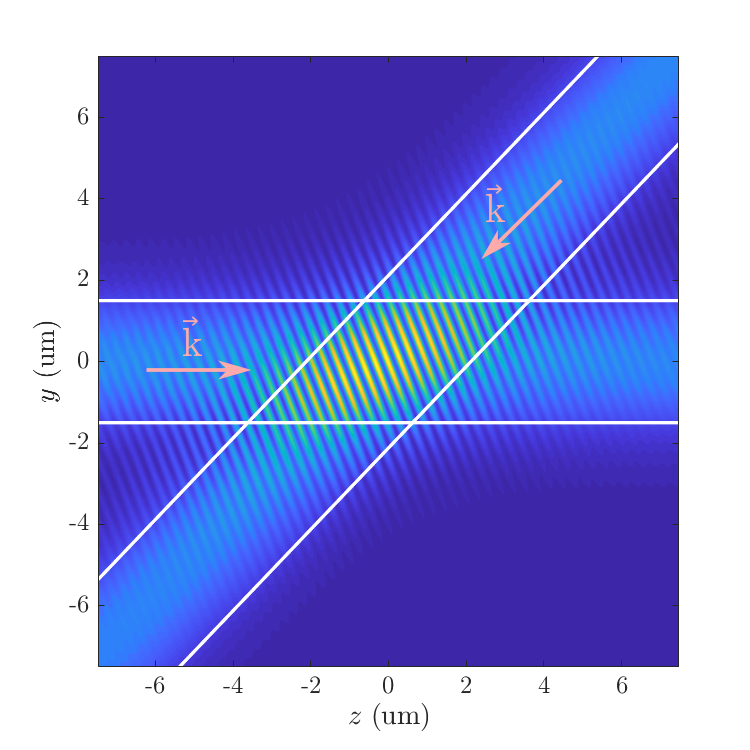}
	\caption{Interference pattern at the cross between two $w$=3 $\mu$m, $h$=300 nm and $h_r$=7 nm waveguides, calculated for light at 720 nm. The waveguides cross at 135 degrees.}
	\label{fringes}
\end{figure}

\begin{figure}
	\centering
	\includegraphics[scale=0.5]{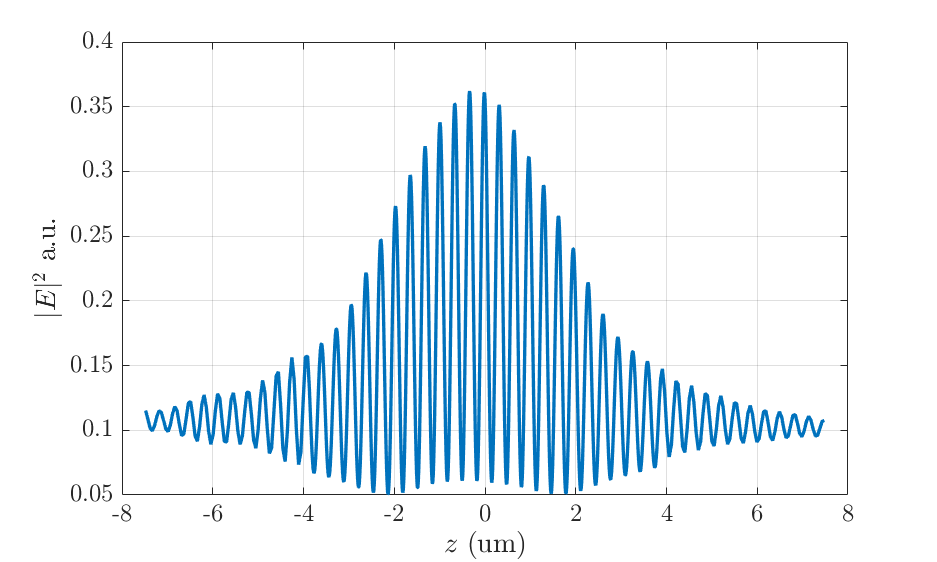}
	\caption{Interference pattern at the cross between two $w$=3 $\mu$m, $h$=300 nm and $h_r$=7 nm waveguides, calculated for light at 720 nm, taken along y=0 line of Fig.15.}
	\label{fringes crossection}
\end{figure}

The other type of beam splitter is based on Bragg diffraction of atoms at an optical lattice formed by interference of optical modes in crossed optical waveguides, which is similar to the waveguide beam splitter based on crossed Gaussian laser beams \cite{13}.

To form an optical lattice in the cross region for such a splitter either red-detuned or blue-detuned modes of the two waveguides should at least partially interfere with each other. For the TE modes of the waveguides such partial interference is possible only when the intersection angle between the waveguides is not equal to 90 degrees. For the normal angle of the intersection between the waveguides the interference is possible only for the TM modes of the waveguides.

Numerical simulations were carried out in order to assess the additional losses due to the intersection of two identical waveguides: $w$=2 $\mu$m, $h$=300 nm $h_r$=15 nm  and  $w$=3 $\mu$m, $h$=300 nm $h_r$=7 nm. To do this, we proceeded in two steps. First, we computed the propagation loss for a straight waveguide. Then, the second rib waveguide was introduced and the propagation loss computed again. The loss in dB due to the intersection was obtained by subtracting the propagation loss of the straight waveguide from the one with intersection. Table 2 shows the losses due to the intersection. Based on the simulations, we conclude that the losses are caused by scattering of light at edges of the cross region, such as that scattered light is lost from further guiding by the waveguides.

\begin{table}
	\begin{center}
	  \caption{\label{jlab1}Losses at the intersection of the waveguides for TE polarisation}
	   \footnotesize
         \begin{tabular}{|c|c|c|c|}
          \hline 
           \multirow{2}{*}{ ${\lambda}$(nm)}  & Angle (${{}^\circ}$) & \multicolumn{2}{c|}{Losses (dB)} \\ 
             \cline{3-4} 
                &  & $w$=2 $\mu$m & $w$=3 $\mu$m  \\ 
                    \hline
                      720  & 45 & 0.03 &0.02 \\ \hline 
                      720 & 90& 0.02 & 0.01 \\ \hline 
				    850 & 45 & 0.03 &0.02\\ \hline 
				    850 & 90 & 0.02 & 0.01\\ \hline 
           \end{tabular} 
       \end{center}
   \end{table}
\normalsize

For the case of 720 nm and 850 nm optical modes of the 3 $\mu$m wide rib waveguide for guiding of rubidium atoms, considered in the previous section, it is preferable to realise interference for the 720 nm mode, because it is producing a smaller period of the optical lattice, which is important for the relatively small dimension of the cross region. The corresponding optical lattice, formed by the intersection 720 nm evanescent waves is shown in Fig.15. A spatial distribution of the light intensity along $y=0$ line is presented in Fig.16. The period of optical lattice of 302.5 nm in that case is determined by the wave vector $k_{eff}=k_0 n_{eff}$, where $k_0$ is the wavevector in vacuum and $n_{eff}$ is the effective index of refraction of the waveguide, which in this case is $n_{eff}$=1.288. To avoid formation of the second optical lattice by crossed 850 nm optical modes it is possible to make their optical frequencies in the two crossed waveguides slightly different, by ~100 MHz or more. Interference of such modes of different frequencies produces a rapidly moving optical lattice, which does not affect the motion of atoms, because of their inertia. This rapidly moving optical lattice will lead only to some modulation of the atomic velocity, which can be made much smaller than the initial velocity spread of the atomic wave packet by proper choose of the frequency difference between the interfering modes. 

Note, that in a case of intersecting waveguides with a single atomic mode, the atoms in such a Bragg splitter can be diffracted only into the single mode of the other waveguide or they are lost from further guiding. A detailed treatment of the waveguide Bragg beam splitter is outside the scope of this paper.

\subsection{Optical losses in bent rib waveguides}

In order to design a final application like a quantum gyroscope based on atomic Sagnac interferometer, a bent waveguide is needed. Here, we assess the losses for propagation around a circle for $w$=2 $\mu$m, $h$=300 nm and $h_r$=15 nm and $w$=3 $\mu$m, $h$=300 nm and $h_r$=7 nm waveguides. Figure \ref{bend_losses} shows the propagation losses as a function of the radius of the circle for both wavelengths for TE polarisation. As expected, losses increase while the radius decreases and are higher at 850 nm than 720 nm. 

\begin{figure}
	\centering
	\includegraphics[scale=0.15]{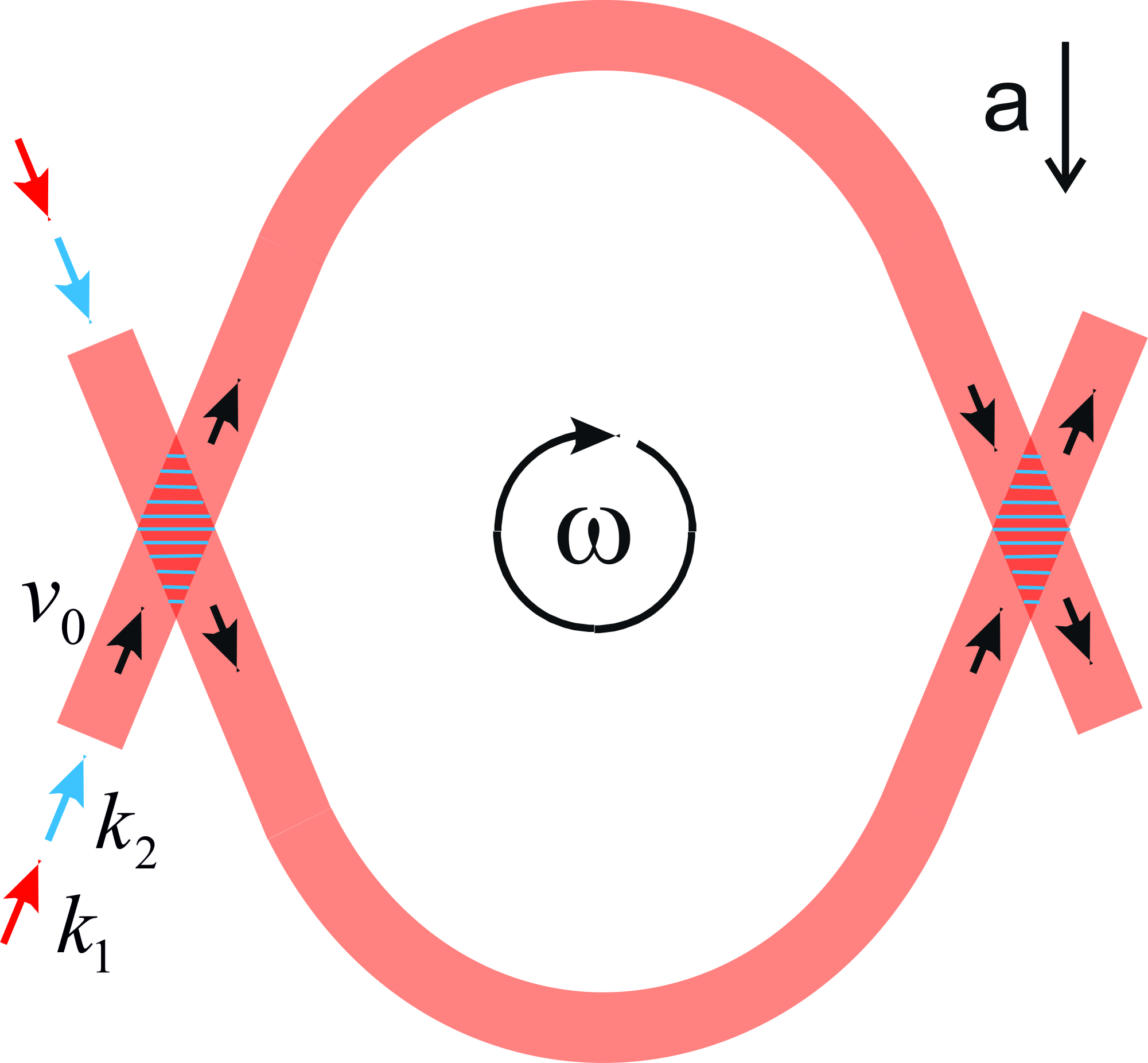}
	\label{}
	\caption{Mach-Zehnder all-optical waveguide interferometer for atoms.}
\end{figure}

\subsection{All-optical waveguide Mach-Zehnder interferometer for atoms}

The Mach-Zehnder waveguide atom interferometer (Fig.17) consists of two bent rib waveguides, which intersect with each other twice at the same angle. Both of these optical waveguides are single modes at different colours, the evanescent waves of which are used for guiding of ultra-cold atoms. The interfering blue-detuned modes produce the optical lattices in the cross regions, which are used as continuous waveguide Bragg beam splitters for matter waves. Ideally both of these beam splitters should have 50:50 splitting ratio. This interferometer can be used for the measurements of both transverse linear accelerations and rotations in the plane of the interferometer. For an area of the interferometer of 1 cm$^2$ and atomic flux $F_{at}=10^6$ at/s and atomic velocity $v=1$ cm/s, the quantum projection noise limited sensitivity of such interferometer to acceleration in plane of the loop is about $S_{acc}\simeq 1\times10^{-10}$ m/s$^2$/$\sqrt{\text{Hz}}$ and its sensitivity to rotation is $S_{rot}=5.7\times10^{-9}$ rad/s/$\sqrt{\text{Hz}}$. Further increase of the sensitivity is possible via an increase of the enclosed area of the interferometers and increase of the atomic flux, while the sensitivity to acceleration can be also increased by reducing the velocity of atoms. It is important to mention, that this interferometer can be operated continuously due to the continuous nature of the waveguide Bragg beam splitter. In a similar way it is possible to design integrated waveguide Michelson and Sagnac interferometers. A Sagnac waveguide interferometer can be based on a ring-like waveguide for atoms, in which two partial matter waves simultaneously propagate in clockwise and anticlockwise directions. The main advantage of such a Sagnac interferometer consists in its sensitivity to rotations, but not to linear accelerations, which makes it an ideal gyroscope. 

\section{Conclusion}

It is shown that suspended rib optical waveguides can be used for coherent quasi-single-mode guiding of ultra-cold atoms by non-resonant evanescent light waves of the two main optical modes of opposite frequency detunings from the main atomic transition. Our modelling shows that in double-evanescent-wave dipole traps, based on waveguides of sub-wavelength width, it is difficult to provide stable transverse trapping of atoms because of small potential depth and high tunnelling rates of atoms in lateral directions. On the other hand, it is much easier to get stable transverse trapping of atoms for waveguides of width from 2 $\mu$m to 3 $\mu$m. Our estimates show that the coherent guiding times of the order of 1 second can be achieved at reasonably small powers of guiding laser light. Continuous waveguide atom interferometers based on such crossed planar all-optical waveguides are discussed.  
Based on these technologies, novel all-optical atom chips could be created for use as quantum inertial sensors, magnetometers, and elements of quantum logic.

\subsection{Acknowledgments}

This work was supported by an ISCF Metrology Fellowship grant provided by the UK government$\textquoteright$s Department for Business, Energy and Industrial Strategy (BEIS).
Many thanks to Richard Hobson, Krzysztof Szymaniec and Geoffrey Barwood for their comments and corrections to the paper.

\end{document}